\documentclass[aps,reprint,article,prx]{revtex4-1}
\usepackage{amsmath}
\usepackage{dcolumn}
\usepackage{graphicx}
\usepackage{float}
\usepackage{ulem}

\begin{document}

\title{No observation of chiral flux current in the topological kagome metal CsV$_{3}$Sb$_{5}$}

\author{Huazhou Li$^{1,*}$, Siyuan Wan$^{1,*}$, Han Li$^1$, Qing Li$^1$, Qiangqiang Gu$^1$, Huan Yang$^{1,\dag}$, Yongkai Li$^{2,3}$, Zhiwei Wang$^{2,3}$, Yugui Yao$^{2,3}$, and Hai-Hu Wen$^{1,\ddag}$}

\affiliation{$^1$ National Laboratory of Solid State Microstructures and Department of Physics, Collaborative Innovation Center of Advanced Microstructures, Nanjing University, Nanjing 210093, China}

\affiliation{$^2$ Key Laboratory of Advanced Optoelectronic Quantum Architecture and Measurement (MOE), School of Physics, Beijing Institute of Technology, Beijing 100081, China}

\affiliation{$^3$ Beijing Key Lab of Nanophotonics and Ultrafine Optoelectronic Systems, Beijing Institute of Technology, Beijing 100081, China}

\begin{abstract}
Compounds with kagome lattice usually host many exotic quantum states, including the quantum spin liquid, non-trivial topological Dirac bands and a strongly renormalized flat band, etc.
Recently an interesting vanadium based kagome family $A$V$_{3}$Sb$_{5}$ ($A$ = K, Rb, or Cs) was discovered, and these materials exhibit multiple interesting properties, including unconventional
saddle-point driving charge density wave (CDW) state, superconductivity, etc. Furthermore, some experiments show anomalous Hall effect which inspires that there might be some chiral flux current states.
Here we report scanning tunneling measurements by using spin-polarized tips. Although we have observed clearly the $2a_0\times2a_0$ CDW and $4a_0$ stripe orders, the well-designed experiments with refined
spin-polarized tips do not reveal any trace of the chiral flux current phase in CsV$_3$Sb$_5$ within the limits of experimental accuracy. No observation of the local magnetic moment in our
experiments may put an upper bound constraint on the magnitude of magnetic moments induced by the possible chiral loop current which has a time-reversal symmetry breaking along $c$-axis in CsV$_{3}$Sb$_{5}$. 
\end{abstract}

\maketitle
\section{Introduction}
The kagome lattice is a planar network composed of hexagons with corner-sharing triangles. Owing to its unique geometry, the antiferromagnetic (AF) long-range order cannot exist due to the spin frustration, resulting in a possible quantum spin liquid state \cite{Sachdev,LeonBalents,spin liquid Science,HanTH.Nature}. Large anomalous Hall effect, which is induced by the spontaneous magnetization, has been observed in some exotic AF states \cite{Tomoya Higo antiferro,Parkin.antiferro} or ferromagnetic semimetals \cite{Nat.Phys.ferro, Nat.Commun.ferro} with the kagome lattice. In addition, materials with a kagome lattice structure can host other quantum states, such as charge density wave (CDW) \cite{Franz.CDW,WangQH}, superconductivity \cite{WenXG.SC,Thomale.SC,WangQH}, topological electronic states \cite{Franz.CDW,Mazin.Dirac,YeL.Dirac,YinJX.flat band,YLChen.Weyl,LiuZ.flat band,KangM.Dirac}, as well as spin density wave (SDW) \cite{LiJX}. Recently, a new family of vanadium based metals $A$V$_{3}$Sb$_{5}$ ($A$ = K, Rb, Cs) has been discovered with the kagome lattice consisting of vanadium atoms \cite{AV3Sb5}. Subsequently, superconductivity, CDW, and topological non-trivial states are observed in these materials \cite{SCCsV3Sb5,SCKV3Sb5,SCRbV3Sb5}. Since then, plenty of studies have been carried out on superconducting properties \cite{Spintriplet,LiSY,Pressure2,Pressure3,Pressure4,GaoHJ,WangZY,YuanHQ,Gaptheory1,DongXL,Gaptheory2,NMR,FengDL}, the topologically non-trivial state \cite{WangZY,LeiHCtopo,ARPES2,ARPES3} and the CDW orders \cite{WangZY,ARPES2,ARPES3,GaoHJ,Hasan KV3Sb5,Zeljikovic CsV3Sb5,CDWtheo1, Optical1,HardX,CDWtheo3,Optical2,CDWtheo4,ARPES1,Zeljikovic KV3Sb5,CDWtheo2,Hasan RbV3Sb5,YaoYugui CsV3Sb5,ARPES4}. In addition, a giant anomalous Hall effect (AHE) has been observed in KV$_{3}$Sb$_{5}$ and CsV$_{3}$Sb$_{5}$ \cite{AHE1,AHE2}, and this effect is detectable below the CDW transition temperature \cite{AHE2}.

Usually, AHE breaks the time-reversal symmetry (TRS), thus a theoretical model with the hypothesis of a chiral flux current phase is proposed to explain the TRS breaking along $c$-axis direction in this family of kagome materials \cite{HuJP}. However, the first muon spin resonance ($\mu$SR) measurement shows the absence of local magnetic moments and long-range magnetic orders in KV$_3$Sb$_5$, and the enhanced signal of the spin relaxation rate below the CDW transition temperature is attributed to the nuclear magnetic moment \cite{uSR1}. Afterwards, it was claimed that the TRS breaking may happen at the magnetic field higher than 3 T, and the signal is dominated by the electronic contribution \cite{uSR2}; while, at a small field lower than 1 T it is still argued as the contribution from the nuclear magnetic moment. Based on a recent $\mu$SR work \cite{uSR3}, the spin relaxation rate gets enhanced and is attributed to a local magnetic field component parallel to the $ab$-plane, but the variation is in the range of $\pm5\%$ along $c$-axis; the TRS breaking is proposed to explain their data based on some kind of models. Therefore, the direct evidence is still lacking for the TRS symmetry breaking along $c$-axis.

Besides, several physical properties do exhibit the breaking of the sixfold symmetry of the crystal structure in the $ab$-plane. A twofold symmetry of $c$-axis resistivity is observed in CsV$_{3}$Sb$_{5}$ with in-plane rotating magnetic field in both the normal and the superconducting states \cite{XiangY}, which suggests the existence of a nematic electronic state and a twofold symmetry of superconducting gap. The nematic electronic state disappears near the CDW transition temperature \cite{XiangY}, and it may be related to the anisotropic in-plane $2a_{0} \times 2a_{0}$ ($2 \times2 $) CDW order \cite{GaoHJ,ARPES2,Hasan KV3Sb5,Zeljikovic KV3Sb5,Hasan RbV3Sb5,YaoYugui CsV3Sb5}. The three-dimensional CDW order is identified as the configuration of $2a_0\times2a_0\times2c_0$ \cite{WangZY,HardX} or $2a_0\times2a_0\times4c_0$ \cite{Wilson3DCDW} by different kinds of experiments. This means that a space shift of $a_0$ emerges along one of three in-plane crystalline $a$-axes for the same characteristic in-plane CDW patterns in neighbored layers, which results in a $\pi$ phase shift between neighbored V-Sb kagome layers. Based on this picture, the phase shift together with the inter-layer coupling between the neighbored layers lower the six-fold symmetry to a twofold one \cite{CDWtheo2,MiaoH2}. The intrinsic chiral anisotropy with the magnetic-field tunability was reported for the $2\times2$ CDW order in $A$V$_{3}$Sb$_{5}$ \cite{Hasan KV3Sb5,Hasan RbV3Sb5}, which was explained as the unconventional chiral CDW in the frustrated kagome lattice. In regions with defects, the CDW order somehow does not show magnetic field response \cite{Hasan KV3Sb5,Hasan RbV3Sb5}. However, the CDW order is argued to be insensitive to the magnetic field even in the defect-free region from another report \cite{Zeljikovic KV3Sb5}, which seems contradicting with previous reports \cite{Hasan KV3Sb5,Hasan RbV3Sb5}. Consequently, the conclusion about both the chiral flux current phase and tunable chiral charge order remains controversial, and they should be examined by more refined experiments, such as using the spin-polarized scanning tunneling microscopy/spectroscopy (SP-STM/STS).

It should be noted that the SP-STM/STS is a very useful tool to investigate magnetic structures in the nanoscale \cite{Wiesendanger1}, and some magnetic orders have been resolved in topological kagome materials by this technique \cite{HasanSPTopo,GaoSPTopo}. Here, we report the measurements by using SP-STM/STS on single crystals of CsV$_3$Sb$_5$. Within our experimental resolution, we have not found any indication of the spin texture corresponding to the proposed chiral flux current phase. Furthermore, we have neither observed any detectable spin textures of the $2\times2$ CDW order, which may rule out the existence of a SDW order.

\section{Methods}
Single crystals of CsV$_{3}$Sb$_{5}$ were synthesized via the self-flux method \cite{AV3Sb5}. The physical properties including those in superconducting and normal states were reported in a previous work \cite{XiangY}. 
The STM/STS measurements were carried out in a scanning tunneling microscope (USM-1300, Unisoku Co., Ltd.) with an ultrahigh vacuum up to $1\times10^{-10}$  torr. A typical lock-in technique is used for the spectrum and 
quasiparticle interference (QPI) mapping  measurements with an ac modulation of 0.5 mV and a frequency of 987.5 Hz. All experimental data were taken at 4.2 K. Here we used a bulk Cr rod to make the tip for STM 
measurements \cite{BulkCrTip}. The spin-polarized Cr tip was electrochemically etched by 3 mol/L NaOH solution with the immersed end covered by a polytetrafluoroethylene tube with a certain length \cite{CrHoffman}. 
After the etching, the Cr tip was transferred into the STM chamber, and the spin polarization feature was characterized on the cleaved surface of single-crystalline Fe$_{1+x}$Te \cite{PeterWahl,WanSY}. 
Then we use the calibrated Cr tips to detect the possibly existing chiral flux current phase in CsV$_{3}$Sb$_{5}$. Although Cr is an AF material, a Cr atom or a few Cr atoms on the apex of the tip may have a 
magnetic moment \cite{WanSY}. Besides, the Cr tip may pick up an iron atom or an iron cluster during the characterization on the surface of Fe$_{1+x}$Te \cite{FeTip}. Both Cr and Fe atom(s) on the apex of the tip can be 
polarized by an external magnetic field. During the characterization of the tip on the surface of Fe$_{1+x}$Te, we can determine that the reverting field of the Cr tip is about $\pm0.7$ T (see Fig.~S1 in the Supplemental 
Material \cite{SI}) by systematic sweeping the magnetic field \cite{Looptip}. In fact, the spin orientations of the tip are never perfectly perpendicular or parallel to the sample surface, there are always the 
in-plane and the out-of-plane components of magnetic moment at the apex of the tip. For example, the Cr tip with the field tuned directions of magnetic moments can be used to detect both the in-plane and the 
out-of-plane components of the sample magnetization \cite{BulkCrTip}. Therefore, we use the calibrated Cr tip to detect the possible magnetization induced by the chiral loop current.

\section{Results}

\subsection{Modelling the topographic image with the chiral flux current phase}

\begin{figure}
\includegraphics[width=8cm]{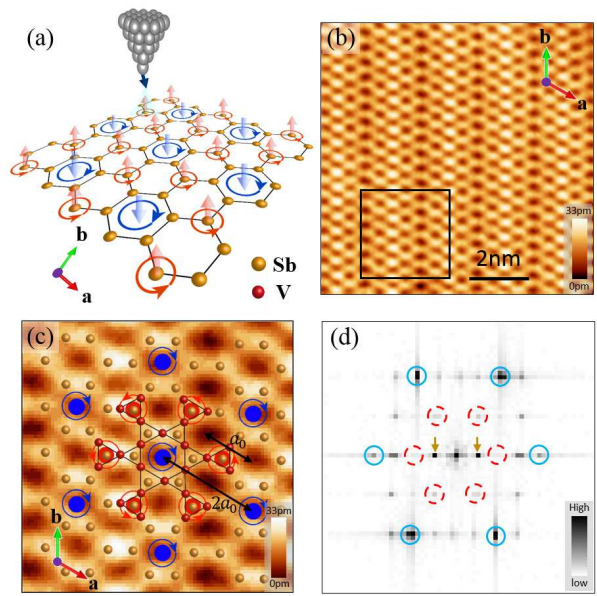}
\caption {(a) Schematic illustration of the spin-polarized STM measurements. (b) Atomically resolved topography of the top Sb surface ($V_\mathrm{set} = 20$ mV, $I_\mathrm{set} = 100$ pA). (c) Enlarged view of the topography in the area of black square in (b). The theoretically proposed hexagonal current loops are shown by blue disks, while the triangular current loops are marked by red disks. (d) Fourier transform of the topography in (b). The Bragg peaks are marked by circles; spots for the $2\times2$ CDW order are marked by dashed circles.
} \label{fig1}
\end{figure}

According to the theoretical proposal of the chiral flux current phase \cite{HuJP}, there are two kinds of current flux loops with opposite circulating directions, and they can produce local magnetic moments with opposite vorticities in the V-Sb layer of $A$V$_{3}$Sb$_{5}$. These flux loops break the TRS in the $c$-axis direction and can have different influences on the tunneling current through the spin-polarized tip with a fixed spin direction. A schematic illustration is given in Fig.~\ref{fig1}(a). Due to the weak van der Waals interaction bridged by Cs and Sb layers, in most cases the easily exposed top layer after the cleavage is composed by Cs or Sb atoms. One can hardly obtain the V-Sb layer as the top layer. In previous STM/STS measurements, the top surface can be Sb, Cs or half Cs surfaces \cite{WangZY,GaoHJ,Zeljikovic CsV3Sb5,WangZY,YaoYugui CsV3Sb5}. Here in our measurements, the commonly obtained surface is the Sb layer, and Fig.~\ref{fig1}(b) shows a typical atomically resolved topography of this kind of surface. The slightly elongated bright spot on the surface represents one pair of Sb atoms. Figure~\ref{fig1}(c) shows an enlarged view of the area marked by a square in Fig.~\ref{fig1}(b). We show together the Sb atoms on the top layer by the yellow spheres as well as those of V atoms underneath by red spheres. It is clear that the Sb atoms in the top layer form a honeycomb like structure. There are dark areas in the center of the hexagons constructed by Sb atoms. The distance between two neighbored hexagonal dark areas equals to the lattice constant $a_{0}$ in the crystal structure. If two kinds of flux loops with opposite magnetic moments existed in the V-Sb layer beneath the top Sb layer, they would locate in one quarter of the total numbers of the hexagonal and triangular vanadium plaquttes.

In Fig.~\ref{fig1}(c), the blue (red) disks denote the positions of hexagonal (triangular) current flux loops based on the theoretical proposal \cite{HuJP}. Here it should be noted that the hexagonal current flux loops appear only in one of every two neighboring lines of dark areas, and on this particular line they only occupy half sites. In this case, the distance between two neighboring hexagonal current flux loops is $2a_{0}$. Figure~\ref{fig1}(d) shows the Fourier transformed (FT) pattern of the topography [Fig.~\ref{fig1}(b)], and one can see several sets of spots corresponding to different periodic structures in the real space besides the Bragg peaks. In the topographic image shown in Fig.~\ref{fig1}(b), one can see clear stripes with the period of $4a_{0}$. The $4a_{0}$ unidirectional order can be clearly observed as spots indicated by arrows in Fig.~\ref{fig1}(d), and they only appear on the horizontal axis in the FT pattern. This result is similar to those from previous reports \cite{GaoHJ,Zeljikovic CsV3Sb5}. The spots corresponding to the $2 \times2 $ CDW order are marked by red dashed circles in FT pattern. The intensities of these spots show a clear anisotropy, and the two spots on the horizontal axis have a relatively weaker intensity comparing to other spots.

\begin{figure*}[htbp]
\centering
\includegraphics[width=13cm]{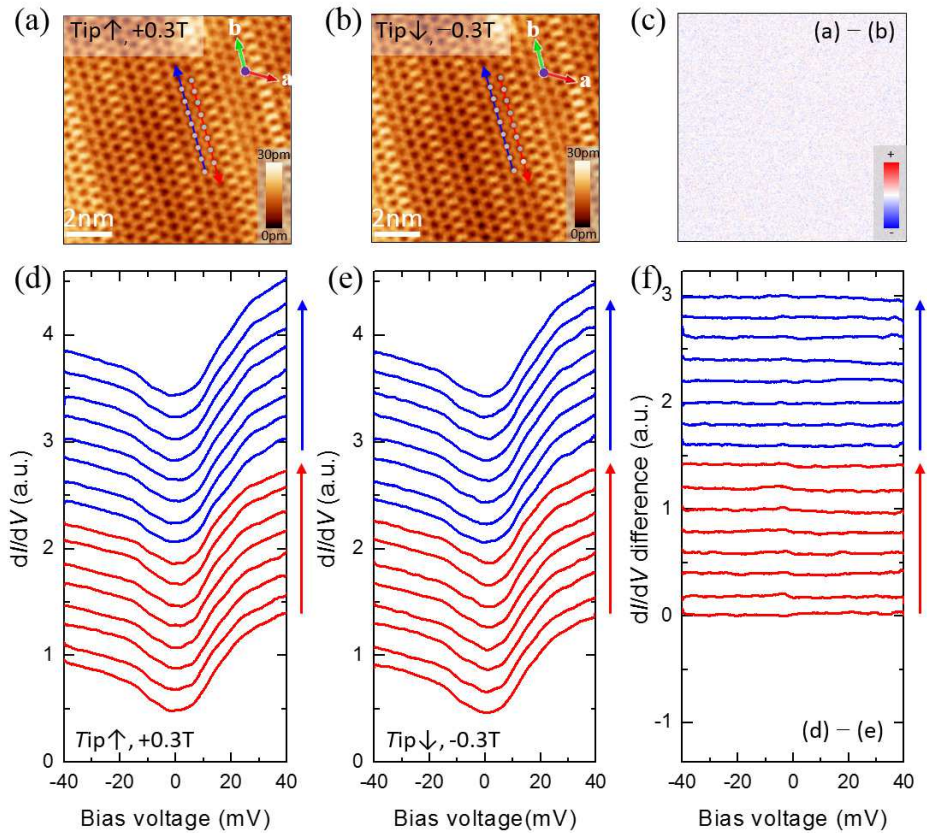}
\caption {(a),(b) Topography of the top Sb surface measured in the same area and under the field of +0.3 and $-0.3$ T; the tip has been polarized by the magnetic fields of +2 (`$\uparrow$') and $-2$ T (`$\downarrow$'), respectively ($V_\mathrm{set} = 30$ mV, $I_\mathrm{set} = 100$ pA). (c) Spin-difference topography between the topographies shown in (a) and (b). (d),(e) Tunneling spectra taken at the centers of hexagonal holes marked by dots in (a) and (b), respectively ($V_\mathrm{set} = 30$ mV, $I_\mathrm{set} = 100$ pA). (f) Spin-difference spectra between the tunneling spectra taken at the same positions but with different tip spin directions and under different magnetic fields. The curves in (d) to (f) are off-set for clarity.
} \label{fig2}
\end{figure*}

\begin{figure*}[htbp]
\centering
\includegraphics[width=18cm]{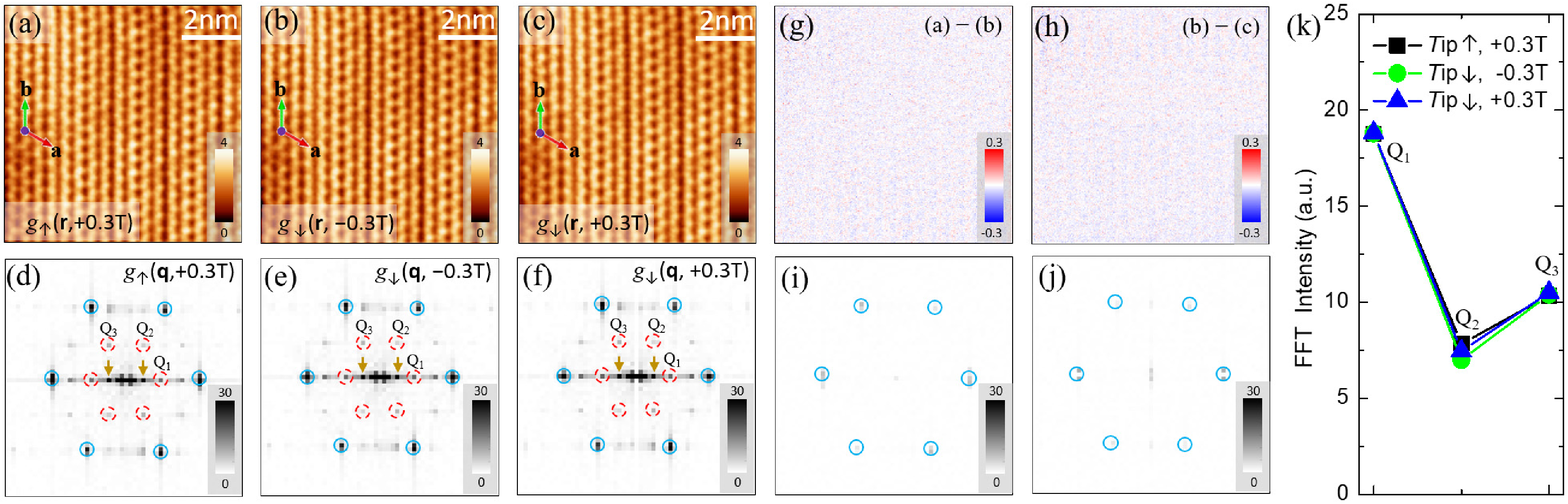}
\caption {(a)-(c) Differential conductance mappings measured at $E = 20$ meV in the same area ($V_\mathrm{set} = 50$ mV, $I_\mathrm{set} = 100$ pA). The magnetic field and the spin polarization of the tips are different in these configurations. (d) to (f) Fourier transform patterns of the differential conductance mappings in (a) to (c), respectively. (g),(h) Spin-difference differential conductance mappings derived by the subtraction of (b) from (a) and (c) from (b), respectively. (i),(j) Fourier transform patterns of spin-difference differential conductance mappings in (g) and (h), respectively. (k) Intensities of scattering spots corresponding to the $2\times2$ CDW order in the FT pattern shown in (d) to (f).
} \label{fig3}
\end{figure*}

\subsection{Topography and tunneling spectrum with spin-polarized tips}
In STM measurements, the obtained topography is a combination of the atomic height and the contribution from the density of states (DOS) \cite{HoffmanReview}. As mentioned above, if the current flux loop can induce a local magnetic moment, they would affect the tunneling current through the spin-polarized tip, which is similar to the situation of investigating magnetic orders in other materials \cite{Wiesendanger1,PeterWahl,WanSY}. For effectively detecting the possible chiral flux current phase with a spin-polarized tip, during the measurements we usually apply a small magnetic field ($\pm 0.3$ T) when the tip polarization is guaranteed. This measure is taken because we need to have a definite TRS breaking state if it really exists. Being different from the exotic AF state \cite{Tomoya Higo antiferro,Parkin.antiferro} or ferromagnetic semimetals \cite{Nat.Phys.ferro,Nat.Commun.ferro} with kagome lattice, the hysteresis behavior is absent in the AHE signals measured in KV$_{3}$Sb$_{5}$ or CsV$_{3}$Sb$_{5}$ \cite{AHE1,AHE2} although a huge AHE signal is observed as a nonlinear part of the Hall resistivity. Since the Hall resistivity is zero at zero magnetic field, the spontaneous magnetization should be negligible, and the $c$-axis symmetry seems to be not broken in these materials at 0 T \cite{AHEReview}. In this point of view, an external magnetic field should be applied to induce the $c$-axis symmetry breaking in CsV$_{3}$Sb$_{5}$. Here in this work, the Cr tip is polarized by magnetic fields of $+2$ or $-2$ T. After magnetizing the tips with magnetic fields of $\pm 2$ T, the external field is decreased to $\pm 0.3$ T for the measurements, respectively. Figures~\ref{fig2}(a) and \ref{fig2}(b) show the topographic images measured at $+0.3$ and $-0.3$ T by using a Cr tip polarized by magnetic fields of $+2$ and $-2$ T, respectively. The symbol of `$\uparrow$' or `$\downarrow$' is used to denote the polarized direction of the tip. Then the spin of the tip should be opposite in these two configurations. Figure~\ref{fig2}(c) shows the difference of the two topographic images measured with opposite spin polarizing directions. However, one cannot see any obvious periodic difference signal in the figure, which suggests that the electronic state in this area does not have detectable spin-sensitive contribution.

Tunneling spectra are measured at positions in the centers of hexagonal dark areas locating along two neighboring lines of these hexagons [red and blue arrowed lines in Figs.~\ref{fig2}(a) and \ref{fig2}(b)]. Among these hexagonal dark areas, half of them should exhibit the hexagonal current flux loops along one of the arrowed lines with a period of $2a_{0}$. The results measured on the neighboring line should not show any spin related signal. Figures~\ref{fig2}(d) and \ref{fig2}(e) show tunneling spectra measured at centers of each hexagonal dark areas along two neighboring routes as highlighted by the arrowed lines. The CDW gapped feature can be observed as two kinks at about $\pm20$ meV in the spectra, which is similar to a previous report \cite{WangZY}. Spin-difference spectra is calculated by the subtraction of two tunneling spectra measured at the same position but with different tip spin orientations and under different magnetic fields, and Fig.~\ref{fig2}(f) shows the corresponding spin-difference spectra derived from those shown in Figs.~\ref{fig2}(d) and \ref{fig2}(e). These spin-difference spectra are almost featureless and quite uniform at positions of hexagonal dark areas, disregard whether the measurement is taken on the line with or without the expected current flux loops. This indicates that our tunneling measurements with spin-polarized tips do not show the signal arising from the expected chiral flux current phase. It should be noted that there is a zero AHE \cite{AHE2} or a zero net magnetic moment \cite{YaoYugui CsV3Sb5} at zero field, and a finite field will induce a finite AHE signal or a finite magnetic moment in the same direction. Based on the model of the chiral flux current \cite{HuJP}, all the magnetic moments induced by the loop currents may reverse when the magnetic field changes its direction in the range from $-1$ to $+1$ T based on the data of the AHE \cite{AHE2}. Concerning Fig.~\ref{fig2}(a),(b), in the scenario mentioned above, it may give no spin contrast signal. However, if the chiral flux current model \cite{HuJP} is correct, the magnetic moment arising from this orbital current should appear only in half of the hexagonal blocks on one of the two neighboring lines. In this case, even the tip polarization is fixed, we should still observe an alternative variation of tunneling current along one of the tracing lines. But this alternative signal cannot be observed in these spectra at any energy in the range from $-40$ to $+40$ meV.

\begin{figure}
\includegraphics[width=8cm]{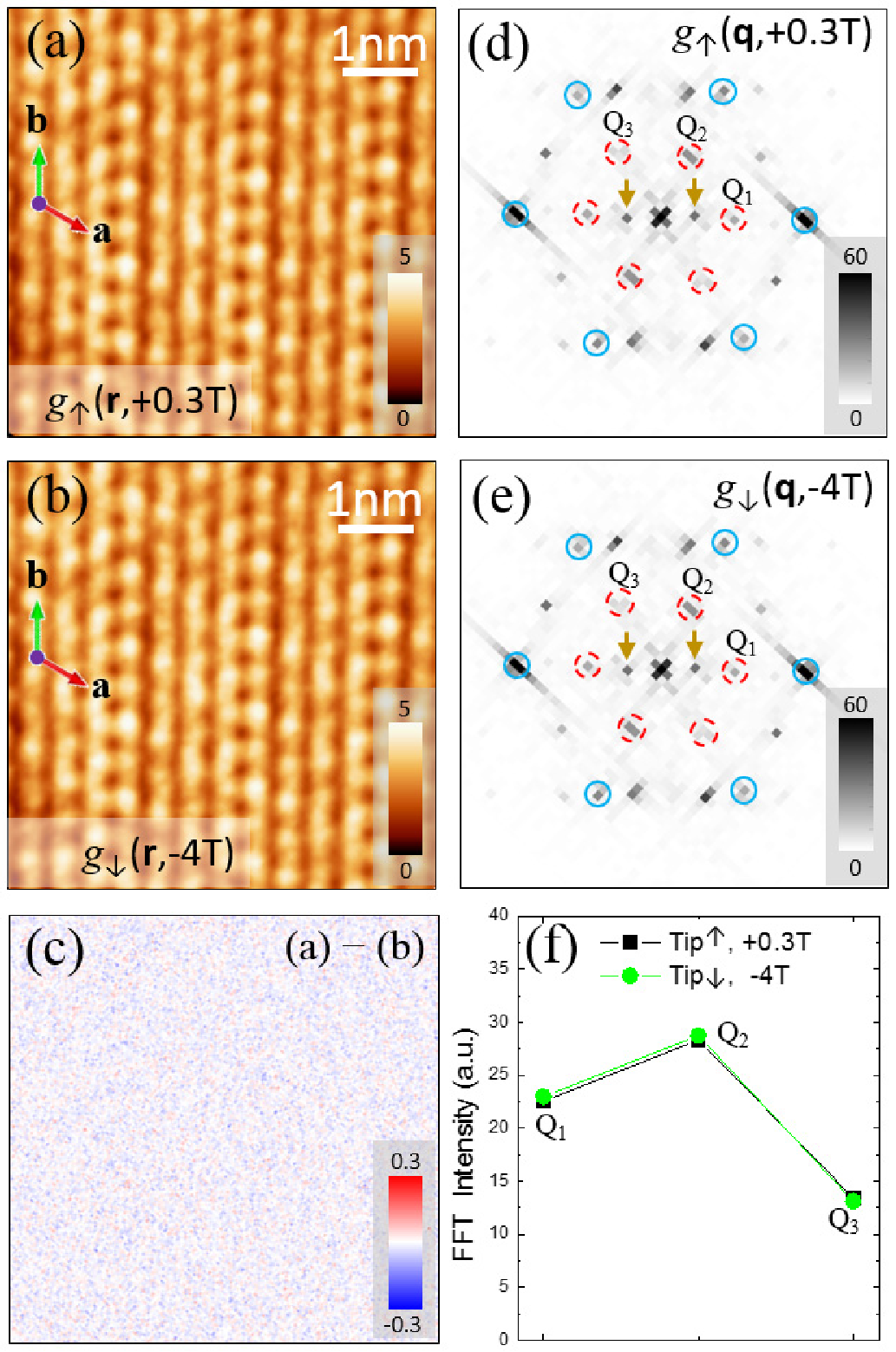}
\caption {(a),(b) Differential conductance mappings measured at $E = 20$ meV in the same area ($V_\mathrm{set} = 30$ mV, $I_\mathrm{set} = 100$ pA). The magnetic field and the spin polarization of the tips are different in these configurations.  (c) Spin-difference differential conductance mappings derived by the subtraction of (b) from (a). (d),(e) Fourier transform patterns of the differential conductance mappings in (a) and (b), respectively. (f) Intensities of the FT patterns of the $2\times2$ CDW order shown in (d) and (e).
} \label{fig4}
\end{figure}

\subsection{Differential conductance mapping using spin-polarized tips }
The differential conductance or the quasiparticle interference (QPI) mapping can provide more direct information of the spatial evolution of DOS. Figures~\ref{fig3}(a)-\ref{fig3}(c) show the differential conductance mappings measured with different tip spin orientations at different magnetic fields. There are also some stripes along $b$-axis direction in these QPI mappings, which is originated from the mixed contribution of the $4a_0$ unidirectional stripe order and the $2\times2$ CDW order. Figures~\ref{fig3}(d)-\ref{fig3}(f) show the corresponding FT patterns of Figs.~\ref{fig3}(a)-\ref{fig3}(c), respectively. The spots of the $2\times2$ CDW order can be seen clearly, and they are marked as Q$_{1}$, Q$_{2}$ and Q$_{3}$ for different wave vectors. Obviously, the intensity of the Q$_{1}$ spots is stronger than the intensities of Q$_{2}$ and Q$_{3}$ spots. When we calculate the spin-difference signal between QPI mappings under different tip polarization directions and magnetic fields, the resultant spin-difference QPI mappings are shown in Figs.~\ref{fig3}(g) and \ref{fig3}(h), indicating negligible signal in this area. The difference QPI mappings exhibit featureless behavior not only for the proposed current flux loops but also for the  $2\times2$ CDW order, and the latter rules out the possibility of attributing the gapped feature to the SDW order. Figures~\ref{fig3}(i) and \ref{fig3}(j) shows the FT patterns of Figs.~\ref{fig3}(g) and \ref{fig3}(h), respectively. One can see that the postulated spots due to the chiral flux current phase are completely absent in the original places of the $2\times2$ CDW order, leaving only very weak signal of the Bragg spots.

Actually the configurations in Fig.~\ref{fig3}(a) or \ref{fig3}(b) are very different from Fig.~\ref{fig3}(c) with respect to the tip polarization and the possible flux current states. In the state shown in Fig.~\ref{fig3}(a) [or \ref{fig3}(b)], we first increased the field to $+2$ T (or $-2$ T), then reduced the field to $+0.3$ T (or $-0.3$ T), thus we believe the tip polarization direction should be the same as the magnetic moment along $c$ axis if the loop current exists. In the case of Fig.~\ref{fig3}(c), we change the field from $-2$ to $+0.3$ T. Here, the final field of $+0.3$ T is not strong enough to revert the polarization direction of the tip because the field is far below the reverting field of $+0.7$ T, but the magnetic moment of the loop current is supposed to be aligned upward. Of course, the real case about whether there is a chiral flux current phase and its reverting field are still unknown, but we believe our specially designed experiments should have set up different configurations of the tip polarization and the magnetic moment direction if this possible chiral current phase exists. To further investigate the possible spin texture of the $2\times2$ CDW order, we show the peak intensities of these spots in Fig.~\ref{fig3}(k) from the FT patterns in Figs.~\ref{fig3}(d)-\ref{fig3}(f). Obviously, the differences of peak intensities are also very small for the spots measured with opposite tip polarization orientations and magnetic field directions. This shows no clear feature of the spin texture of the chiral flux current phase with the resolution of our technique.

It should be noted that the chiral charge order was claimed by observing the variation of the intensities of some FT-spots corresponding to the $2\times2$ CDW order by applying a magnetic field of 2 T in KV$_{3}$Sb$_{5}$ \cite{Hasan KV3Sb5} or 3 T in RbV$_{3}$Sb$_{5}$ \cite{Hasan RbV3Sb5}. One may argue that the unobserved spin texture corresponding to the possible chiral charge order mentioned above in our experiments may be because the applied magnetic field of 0.3 T is too small, so a large magnetic field of 4 T is applied to investigate the spin-difference signal. Obviously, a higher magnetic field is supposed to be more effective to tune the direction of the magnetic moment arising from the current flux loops or the chiral charge order. Figure~\ref{fig4} shows the related results measured under magnetic fields of $-4$ and $+0.3$ T. The tip is polarized by the magnetic field of $-4$ and $+2$ T, respectively, in order to have the opposite directions of the tip spin. And the QPI data are recorded at $-4$ and $+0.3$ T in order to induce local moments with different orientations and magnitudes. The magnetic field of 4 T would be strong enough to induce a visible difference among the related FT-QPI spots if the difference were attributed to the chiral charge order \cite{Hasan KV3Sb5,Hasan RbV3Sb5}. From our data shown in Fig.~\ref{fig4}(c), one can see that, however, there is no obvious feature in the spin-difference QPI mapping. In addition, the spot intensities of the FT-QPI patterns depicted in Fig.~\ref{fig4}(f) show also negligible difference. We also measured the differential conductance mappings at different energies and under magnetic fields of $\pm3$ T by using a tungsten tip in a defect-free area, and the result is shown in Fig. S2 in Supplemental Material \cite{SI}. Based on this set of data measured at $\pm3$ T, one can see the anisotropic intensities of CDW spots instead of the chiral anisotropy with the magnetic-field tunability.

\section{Discussion}\label{discussion}
By doing SP-STM/STS measurements in CsV$_{3}$Sb$_{5}$, we have not observed any trace of the chiral flux current phase or the tunable chiral charge order in our experiment. One most possible reason would be that the resolution of our technique may not be high enough to resolve this phase. However, we must mention that, using the same technique we have successfully visualized an emergent incommensurate AF order in the nearby region of magnetic Fe impurities embedded in the optimally doped Bi$_{2}$Sr$_{2}$CaCu$_{2}$O$_{8+\delta}$ and the AF order in Fe$_{1+x}$Te \cite{WanSY}, which proves the ability to detect the local magnetic moment by a spin-polarized Cr tip. According to a previous work \cite{FeTip}, the authors successfully observed an AF order by using a Cr tip in the STM measurements on the lightly doped Sr$_2$IrO$_4$. The magnetic moment at each Ir atom is about $0.2\mu_\mathrm{B}$ (with $\mu_\mathrm{B}$ the Bohr magneton) in an undoped sample \cite{MomentSr2IrO4}, and thus we speculate that the lower bound resolution of the Cr tip is better than $0.2\mu_\mathrm{B}$. Here in CsV$_{3}$Sb$_{5}$, the chiral flux loop generally can produce a magnetic moment in the center of the loop of the orbital current. However, since it is not clear in the theoretical work \cite{HuJP} how large the magnitude is for the magnetic moment produced by this possible loop current, it is difficult for us to make a definite judgement. The absence of spin-difference signal in CsV$_{3}$Sb$_{5}$ may be the intrinsic feature of this material. Here in this family of materials, the chiral flux current phase is proposed to explain the TRS breaking along $c$-axis direction \cite{HuJP} and the observation of the anomalous Hall effect \cite{AHE1,AHE2}. Being different from the spin magnetism, the chiral flux loop generally has a tiny magnetic moment. It has not been established that spin-polarized STM can be an effective methodology to detect such a novel phase. More spin-sensitive proofs like $\mu$SR can be a more direct technique to address this question. However, the $\mu$SR experiments show that the enhanced signal of the spin relaxation rate below the CDW temperature is very small along $c$-axis when the magnetic field is lower than 1 T \cite{uSR1,uSR2,uSR3}, and the direct evidence of the TRS breaking along $c$-axis is still lacking. Hence, the absence of spin-difference signal from our experimental data measured at $\pm0.3$ T is understandable. In addition, we have not observed the tunable intensity of the $2\times2$ CDW spots at different magnetic fields up to 4 T with different directions of the tip spin, which is consistent with a recent report \cite{Zeljikovic KV3Sb5}. It seems that the $2\times2$ CDW order does not have the spin texture corresponding to the proposed chiral charge order from our SP-STM/STS measurements. From a recent $\mu$SR measurement \cite{uSR2}, a factor of six enhancement of the spin relaxation rate at 8 T compared to zero-field is observed in KV$_{3}$Sb$_{5}$, this field is actually much higher than the saturation field ($\sim$1 T) of the AHE \cite{AHE2}, which poses a question that whether the enhanced spin relaxation signal under a high field is really resulted from the chiral charge order. Within the limits of experimental accuracy, we can conclude that the magnetic moment arising from the chiral loop current is smaller than $0.2\mu_\mathrm{B}$ if it exists. Certainly this remains still as an unsettled issue, more refined experiments are highly desired.

\section{Conclusion}\label{Conclusion}
By using spin-polarized tips, we have carried out careful measurements of scanning tunneling microscopy and spectroscopy on the recently discovered kagome metal CsV$_{3}$Sb$_{5}$. Although our data reveal clear evidence of the $2\times2$ CDW and $4a_0$ stripe orders, but refined analyses give no trace for the existence of the proposed chiral charge order or chiral current flux phase. This conclusion is drawn from the results of many thoughtful measurements and analyses: from both the absence of distinction between the spectra with and without the theoretically predicted local magnetic moments on neighboring hexagons of vanadium atoms, and negligible intensity difference of the Fourier transform spots under different magnetic fields with different tip polarization directions. However, no observation of such signal here may not indicate the absence of the possible chiral orbital current phase due to the limited resolution of our experiments. Our results put a constraint on the upper bound signal of magnetic moments arising from this time reversal symmetry breaking phase.

\begin{acknowledgments}
We appreciate the useful message concerning the technique of spin-polarized tunneling measurement from Roland Wiesendanger. This work was supported by National Natural Science Foundation of China (Grants No. 11927809, No. 11974171, No. 12061131001, No. 92065109, No. 11734003, and No. 11904294), National Key R\&D Program of China (Grant No. 2020YFA0308800), Strategic Priority Research Program (B) of Chinese Academy of Sciences (Grant No. XDB25000000), Beijing Natural Science Foundation (Grant No. Z190006), and Beijing Institute of Technology Research Fund Program for Young Scholars (Grant No. 3180012222011). Z.W. would like to thank Micronano Center at BIT for the assistance of sample characterization.
\end{acknowledgments}

$^{*}$ These authors contributed equally to this work.

$^{\dag}$ huanyang@nju.edu.cn

$^{\ddag}$ hhwen@nju.edu.cn

\end{document}